\documentclass[aps,preprint,floats,epsf,epsfig,nofootinbib,letter]{revtex4}

\usepackage{epsfig}
\usepackage{graphicx}
\usepackage{dcolumn}
\usepackage{bm}
\usepackage{subfigure}

\def\DESepsf(#1 width #2){\epsfxsize=#2 \epsfbox{#1}}

\begin{document}
\def\be{\begin{eqnarray}}
\def\en{\end{eqnarray}}
\def\non{\nonumber}
\def\la{\langle}
\def\ra{\rangle}
\def\Br{{\mathcal B}}
\def\BB{{{\cal B} \overline {\cal B}}}
\def\BD{{{\cal B} \overline {\cal D}}}
\def\DB{{{\cal D} \overline {\cal B}}}
\def\DD{{{\cal D} \overline {\cal D}}}
\def\sq{\sqrt}
\def\ov{\overline}


\title{On the smallness of Tree-dominated Charmless Two-body Baryonic $B$ Decay Rates}

\author{Hai-Yang Cheng$^{1}$, Chun-Khiang Chua$^{2}$}
\affiliation{$^1$ Institute of Physics, Academia Sinica,
Taipei, Taiwan 115, Republic of China,
$^2$ Department of Physics and Center for High Energy Physics,
Chung Yuan Christian University,
Chung-Li, Taiwan 320, Republic of China}

\date{\today}

\begin{abstract}
The long awaited baryonic $B$ decay $\overline B^0\to p\bar p$ was recently observed by LHCb with a branching fraction of order $10^{-8}$. All the earlier model predictions are too large compared with experiment. In this work, we point out that for a given tree operator $O_i$, the contribution from its Fiertz transformed operator, an effect often missed in the literature, tends to cancel the internal $W$-emission amplitude induced from $O_i$.
The wave function of low-lying baryons are symmetric in momenta and the quark flavor with the same chirality, but antisymmetric in color indices.  Using these symmetry properties and the chiral structure of weak interactions, we find that half of the Feynman diagrams responsible for internal $W$-emission cancel.
Since this feature holds in the charmless modes but not in the charmful ones, we advocate that the partial cancellation accounts for the smallness of the tree-dominated charmless two-body baryonic $B$ decays.
This also explains why most previous model calculations predicted too large rates as the above consideration was not taken into account. Finally, we emphasize that, contrary to the claim in the literature, the internal $W$-emission tree amplitude should be proportional to the Wilson coefficient $c_1+c_2$ rather than $c_1-c_2$.

\end{abstract}


\maketitle

\section{Introduction}
A unique feature of hadronic $B$ decays is that the $B$ meson is heavy enough to allow a baryon-antibaryon pair production in the final state (for a review of baryonic $B$ decays, see \cite{Bevan,Cheng:2006}).
Naively, it is tempting to expect a large fraction of baryonic decays to
proceed via two-body decay channels due to the larger phase space available for them.
However,  it has been found experimentally that decays of
$B$ mesons to just a baryon and an antibaryon are rare and have
smaller branching fractions than the three-body ones, for example, $\Br(\ov B^0\to p\bar p)\ll\Br(B^-\to p\bar p K^-)$ and $\Br(B^-\to\Lambda\bar p)\ll \Br(\ov B^0\to \Lambda \bar p\pi^+)$ \cite{PDG}.
The first two-body baryonic $B$ decay observed was $\ov B^0\to\Lambda_c^+\bar p$ \cite{Belle:Lamcp}. Subsequently, $B$ mesons decaying
to two charmed baryons, e.g. $B^+\to\bar\Xi_c^0\Lambda_c^+$, were observed with larger rates  \cite{Chistov:2005zb}. No charmless two-body baryonic $B$ decays have
been observed at $B$ factories and the upper limit has
been pushed to the $10^{-7}$ level \cite{PDG}. For example, the most stringent limit on the two-body charmless baryonic decay was set by Belle:  $\Br(\ov B^0\to p\bar p)<1.1\times 10^{-7}$ \cite{Belle:pp}. Very recently, the LHCb collaboration has presented the first
evidence of this mode with the branching fraction $(1.47^{+0.62+0.35}_{-0.51-0.14})\times 10^{-8}$ \cite{LHCb:pp}.

There exist several theoretical models for describing
$B$ decays into two baryons: the pole model \cite{Jarfi,Cheng:2002,Deshpande:1987nc},
the diquark model \cite{Ball,Chang:2001jt} and the QCD sum rule analysis
\cite{Chernyak}. The predictions of
these models for some selected charmless baryonic $B$ decays are listed in the Table II of  \cite{Cheng:2002}. Evidently, many of the earlier model
predictions are too large compared with experiment. For example, the prediction of $\Br(\ov B^0\to p\bar p)$ ranges from $2.7\times 10^{-5}$ \cite{Ball} to $1.1\times 10^{-7}$ \cite{Cheng:2002}.
Hence, most of the previous theoretical predictions are not trustworthy. Presumably a reliable prediction based on pQCD can be
made as the energy release in charmless two-body decay is
very large, justifying the use of pQCD \cite{Cheng:2002}. This approach has been successfully
applied to $\ov B^0\to\Lambda_c^+\bar p$ \cite{He:2006}. The pQCD calculation for charmless modes such as $p\bar p$ and $\Lambda \bar p$
has not yet been carried out.

Using the long awaited $\ov B^0\to p\bar p$ data from LHCb and considering the topological approach together with the chirality structure of weak interactions \cite{Chua:2003it}, one of us (C.K.C.) was able to extract information on topological amplitudes, estimate the penguin to tree amplitude ratio and predict the rates of all other low-lying octet and decuplet modes in the heavy quark limit \cite{Chua:2013zga}.

Even before the LHCb measurement of $\ov B^0\to p\bar p$, it has been argued that its branching fraction is most likely of order $10^{-8}$ \cite{Cheng:2006,Bevan}. This charmless decay  is suppressed relative
to  $\bar B^0\to\Lambda_c^+\bar p$ by the Cabibbo-Kobayashi-Maskawa (CKM) matrix elements $|V_{ub}/V_{cb}|^2$ and is subject to a possible dynamical suppression
\be
\Br(\overline B^0\to p\bar p) &=& \Br(\overline B^0\to\Lambda_c^+\bar p)|V_{ub}/V_{cb}|^2\times f_{dyn} \non \\
&\sim& 2\times 10^{-7}\times f_{dyn}.
\en
A similar relation holds for charmful modes where
the CKM angles for $\Xi_c\bar\Lambda_c$ and $\Lambda_c\bar p$ have
the same magnitudes except for a sign difference
\be
 \Br(\ov B^0\to\Lambda_c^+\bar p)=\Br(\ov
 B^0\to\Xi_c^+\bar\Lambda_c^-)\times f'_{dyn}.
 \en
Experimental measurements \cite{PDG} indicate that the dynamical suppression effect $f'_{dyn}$ is of order $10^{-2}$. This suppression can be understood
from the observation that no hard gluon is
needed to produce the energetic $\Xi_c\ov\Lambda_c$ pair in $B$ decays, while two gluons are needed to produce an
energetic anti-proton in the decay $\ov B^0\to\Lambda_c^+\bar p$. Therefore, the latter process is suppressed relative to the former due to a dynamical suppression $f'_{dyn}\sim {\cal O}(\alpha_s^2)\sim 10^{-2}$ \cite{Cheng:2005vd}.
In the absence of dynamical suppression $f_{dyn}$, the predicted
branching fraction for two-body charmless decays will be of order $10^{-7}$.
If the dynamical suppression is of order $10^{-2}$ similar to that of $\Lambda_c\bar p$ relative
to $\Xi_c\bar\Lambda_c$, then it will become of order $10^{-9}$ and thus beyond
the reach even of super flavor factories. In reality, the
branching fraction is most likely of order $10^{-8}$, between
the extreme cases of $10^{-7}$ and $10^{-9}$.

Since at least two hard gluons are needed in both $\ov B^0\to\Lambda_c^+\bar p$ and $\ov B^0\to p\bar p$ decays,~\footnote{Note that the center of mass momentum for $p\bar p$, $\Lambda_c^+\bar p$ and $\Xi^0_c\bar\Lambda^-_c$ final states is $2.467$, $2.021$ and $1.144$~GeV, respectively~\cite{PDG}. The energy released in the first two modes is much larger than the third one.}
one may wonder where is the underlying source for the dynamical suppression $f_{dyn}$ which is presumably of order $10^{-1}$. In this work, we shall point out that for a given tree operator $O_i$, the effect from its Fiertz transformed operator, a contribution often missed in the literature, tends to cancel the amplitude induced from $O_i$. As a consequence, the smallness of tree-dominated charmless two-body baryonic $B$ decays follows from partial cancellation.

This work is organized as follows. The aforementioned argument for the smallness of tree-dominated charmless two-body baryonic $B$ decays is spelled out in details in Sec. II. In particular we show explicitly that half of Feynman diagrams cancel. Implications of our results are discussed in Sec. III. Sec. IV presents our conclusions.


\section{Tree-dominated two-body baryonic $B$ decay}

The effective weak Hamiltonian for charmless $B$ decays is~\cite{Buras}
\be
H_{\rm eff}=\frac{G_f}{\sq2}
                   \Big\{\sum_{r=u,c} V_{qb}V^*_{uq}[c_1 O^r_1+c_2 O^r_2]
                         -V_{tb} V^*_{tq}\sum_{i=3}^{10} c_i O_i\Big\}+{\rm H.c.},
\label{eq:H_eff1}
\en
where $q=d,s$, and
\be
&& O^r_1=(\bar r_\alpha b_\alpha)_{V-A}(\bar q_\beta r_\beta)_{V-A},
\qquad\qquad
O^r_2=(\bar r_\beta b_\alpha)_{V-A}(\bar q_\alpha r_\beta)_{V-A},
\non\\
&& O_{3(5)}=(\bar q b)_{V-A}\sum_{q'}(\bar q' q')_{V\mp A},
\qquad\quad
O_{4(6)}=(\bar q_\alpha b_\beta)_{V-A}\sum_{q'}(\bar q_\beta' q_\alpha')_{V\mp A},
\non\\
&& O_{7(9)}=\frac{3}{2}(\bar q b)_{V-A}\sum_{q'}e_{q'}(\bar q' q')_{V\pm A},
\quad
O_{8(10)}=\frac{3}{2}(\bar q_\alpha b_\beta)_{V-A}\sum_{q'}e_{q'}(\bar q_\beta' q_\alpha')_{V\pm A},
\label{eq:H_eff2}
\en
with $O_{3-6}$ being the QCD penguin operators, $O_{7-10}$ the electroweak penguin operators, and $(\bar q' q)_{V\pm A}\equiv \bar q'\gamma_\mu(1\pm\gamma_5)q$.
The spin-flavor wave
function of a left-handed (helicity$=-\frac{1}{2}$) low-lying octet or decuplet baryon can be
expressed as (see, for example, \cite{Lepage:1979za})
\begin{equation}
|{\cal B}\,;\downarrow\rangle\sim \frac{1}{\sqrt3}(|{\cal
B}\,;\downarrow\uparrow\downarrow\rangle
                +|{\cal B}\,;\downarrow\downarrow\uparrow\rangle
                +|{\cal B}\,;\uparrow\downarrow\downarrow\rangle),
\end{equation}
i.e. composed of 13-, 12- and 23-symmetric terms, respectively.
For ${\cal B}=\Delta^{++,+}, p$, we have
\begin{eqnarray}
|\Delta^{++};\downarrow\uparrow\downarrow\rangle&=&u(1)u(2)u(3)|\downarrow\uparrow\downarrow\rangle,\qquad\qquad
\nonumber\\
|\Delta^{+};\downarrow\downarrow\uparrow\rangle&=&
\frac{1}{\sqrt3}[u(1)u(2)d(3)+u(1)d(2)u(3)+d(1)u(2)u(3)]|\downarrow\downarrow\uparrow\rangle,
\nonumber\\
 |p\,;\downarrow\downarrow\uparrow\rangle&=&
\left[\frac{d(1)u(2)+u(1)d(2)}{\sqrt6} u(3)
 -\sqrt{\frac{2}{3}} u(1)u(2)d(3)\right]
|\downarrow\downarrow\uparrow\rangle,
\end{eqnarray}
for the corresponding $|{\cal
B}\,;\downarrow\uparrow\downarrow\rangle$ parts, while the 12- and
23-symmetric parts can be obtained by permutation.

The quark diagrams for two-body baryonic $B$ decays involve the internal $W$-emission tree diagram for $b \to  c(u)$, the penguin loop diagram for
$b \to s(d)$ transition, $W$-exchange for the neutral
$B$ meson, and $W$-annihilation for the charged
$B$. As for mesonic $B$ decays, $W$-exchange and $W$-annihilation are expected to be helicity suppressed which can be understood in the same way as for leptonic decays.
Therefore, the main contributions to two-body
baryonic $B$ decay $B\to \BB'$ are due to either the internal
$W$-emission diagram or the penguin diagram.

\begin{figure}[t]
\centering
 \subfigure[]{
     \includegraphics[width=0.5\textwidth,natwidth=610,natheight=642]{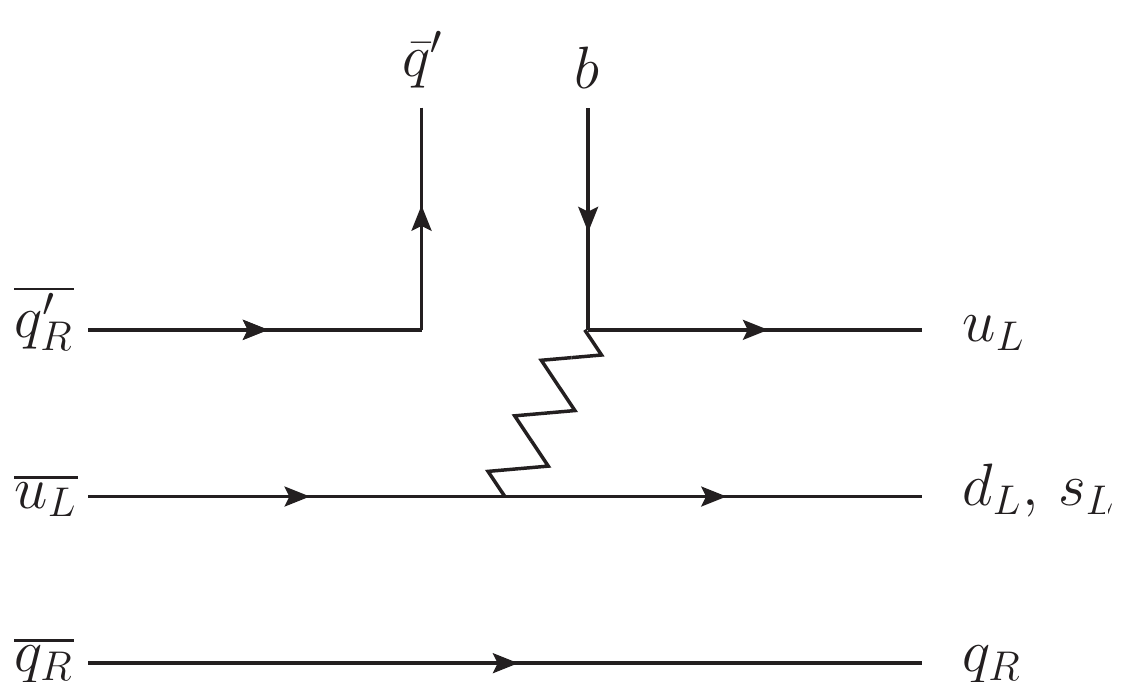}
}\subfigure[]{
    \includegraphics[width=0.5\textwidth,natwidth=610,natheight=642]{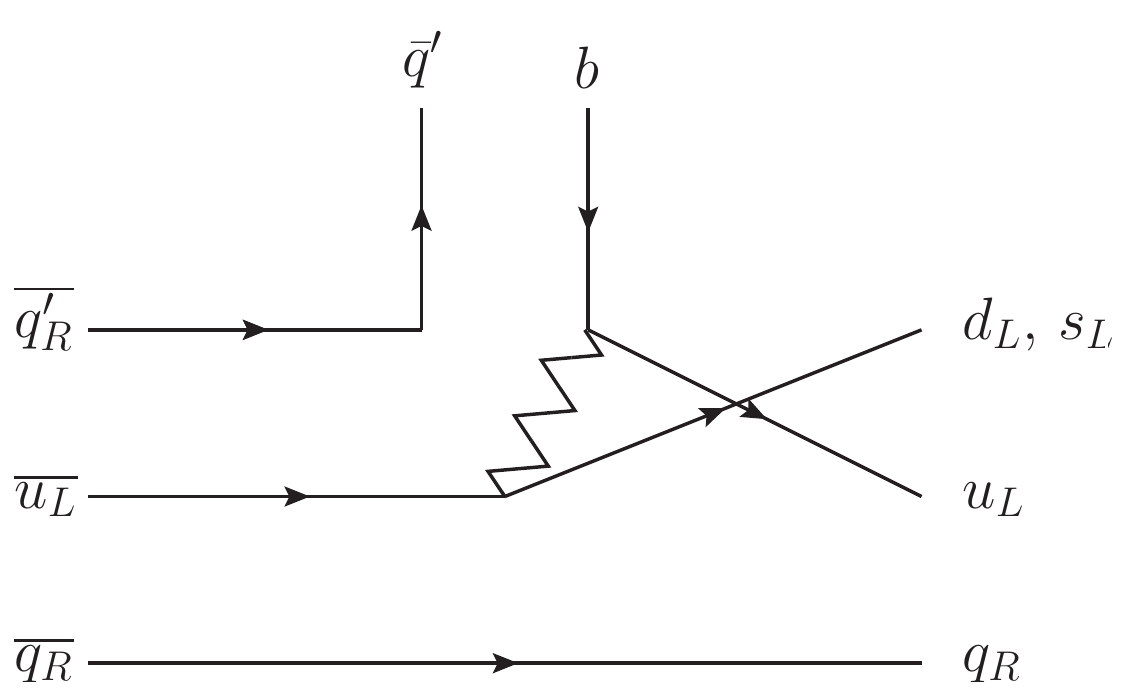}
}
\caption{Tree diagrams in the asymptotic limit. }
\label{fig:eTeT1}
\end{figure}

Two internal $W$-emission diagrams induced by the tree operator $O_1$ in the heavy quark limit are exhibited in Fig.~\ref{fig:eTeT1}.
Intuitively, it is expected that the second diagram will cancel the first one.
The argument goes as follows. After Fiertz reorder, the second diagram can be brought into the first one except for three differences: switching the flavor and momenta of $u_L$ and $d_L (s_L)$ of the final state baryon, and also switching the color quantum numbers of these two quarks in Fig. 1(a).
Since the baryon wave function is symmetric in the flavor and momenta of $u_L$ and $d_L(s_L)$ (see Eqs. (5) and (6)) and antisymmetric in color indices, the second amplitude [Fig. 1(b)] is opposite in sign to the first amplitude [Fig. 1(a)].
As we shall see shortly below, the realistic process of charmless baryonic decays involves at least two hard gluons.
Although the QCD interaction is chiral- and color-conserving, adding gluons will redistribute colors and momenta and change the Dirac structure. Therefore, we need to explicitly check if the above feature still holds.

\begin{figure}[t!]
\centering
 \subfigure[]{
     \includegraphics[width=0.33\textwidth,natwidth=610,natheight=642]{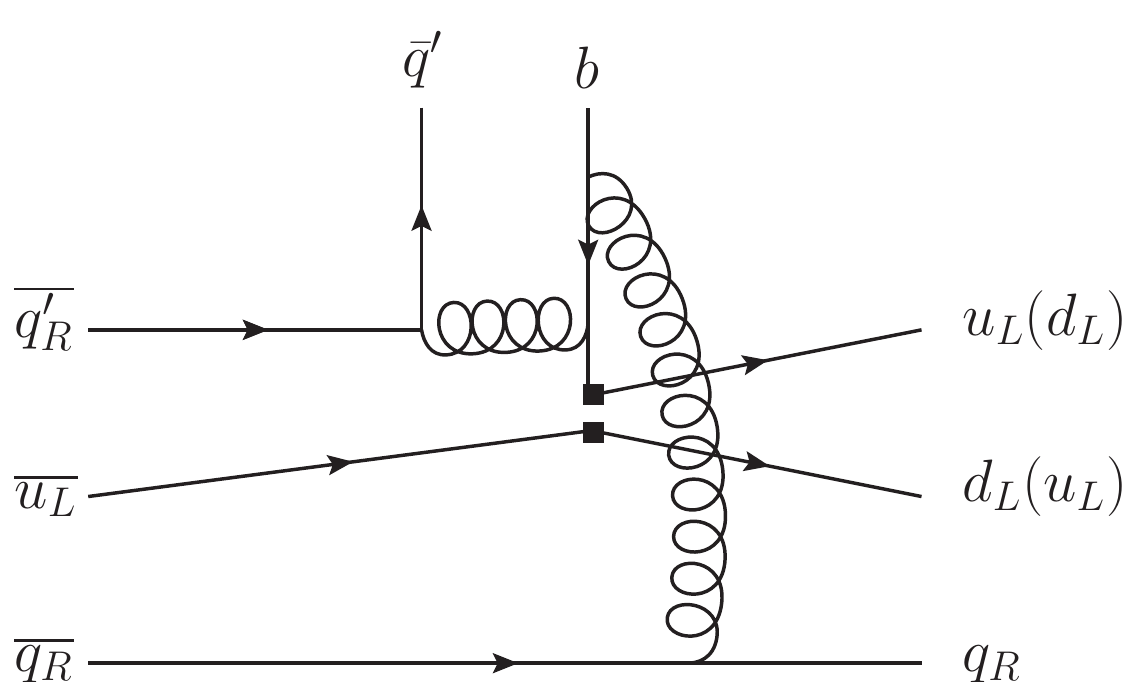}
}\subfigure[]{
    \includegraphics[width=0.33\textwidth,natwidth=610,natheight=642]{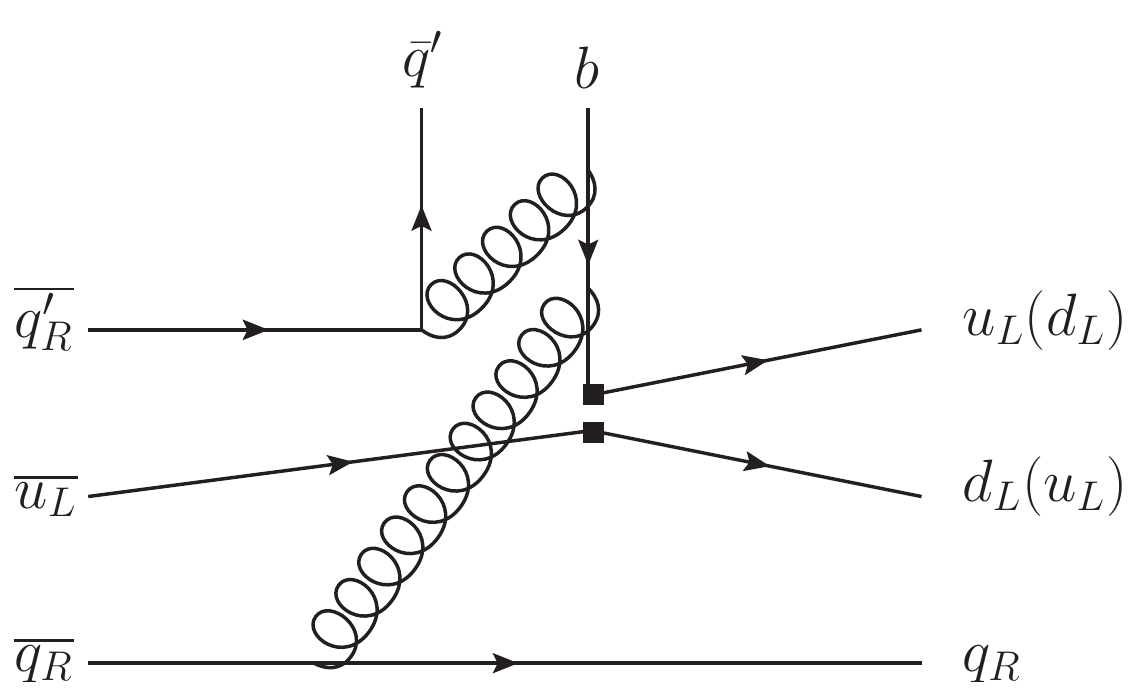}
}\subfigure[]{
     \includegraphics[width=0.33\textwidth,natwidth=610,natheight=642]{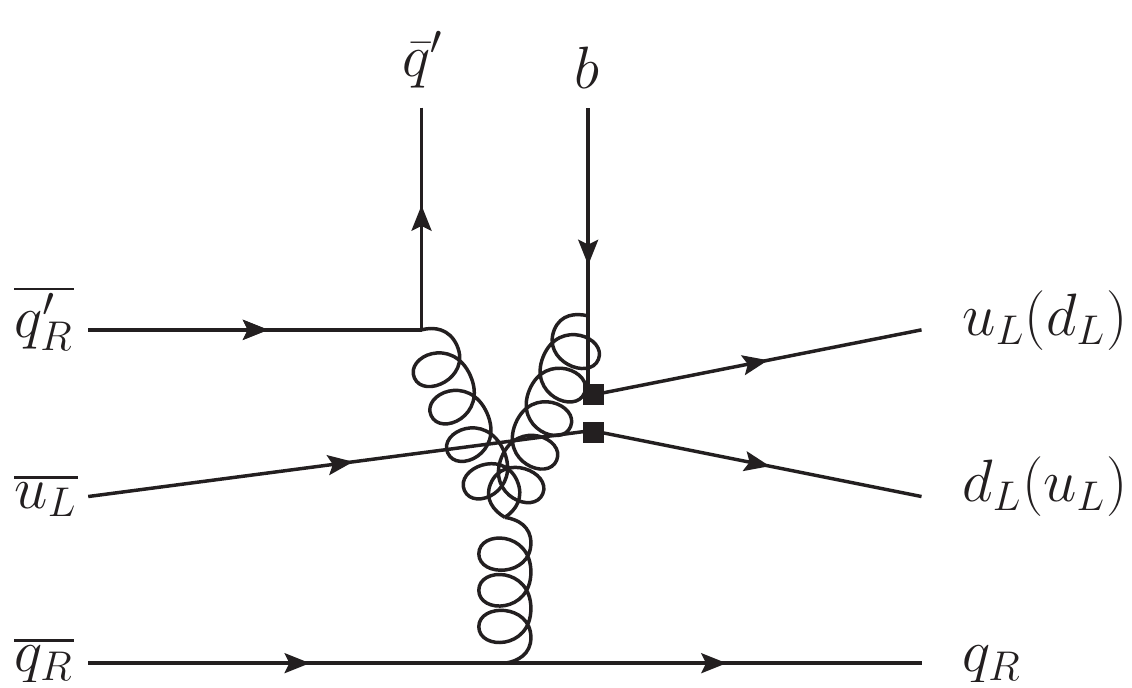}
}
\\
\subfigure[]{
     \includegraphics[width=0.33\textwidth,natwidth=610,natheight=642]{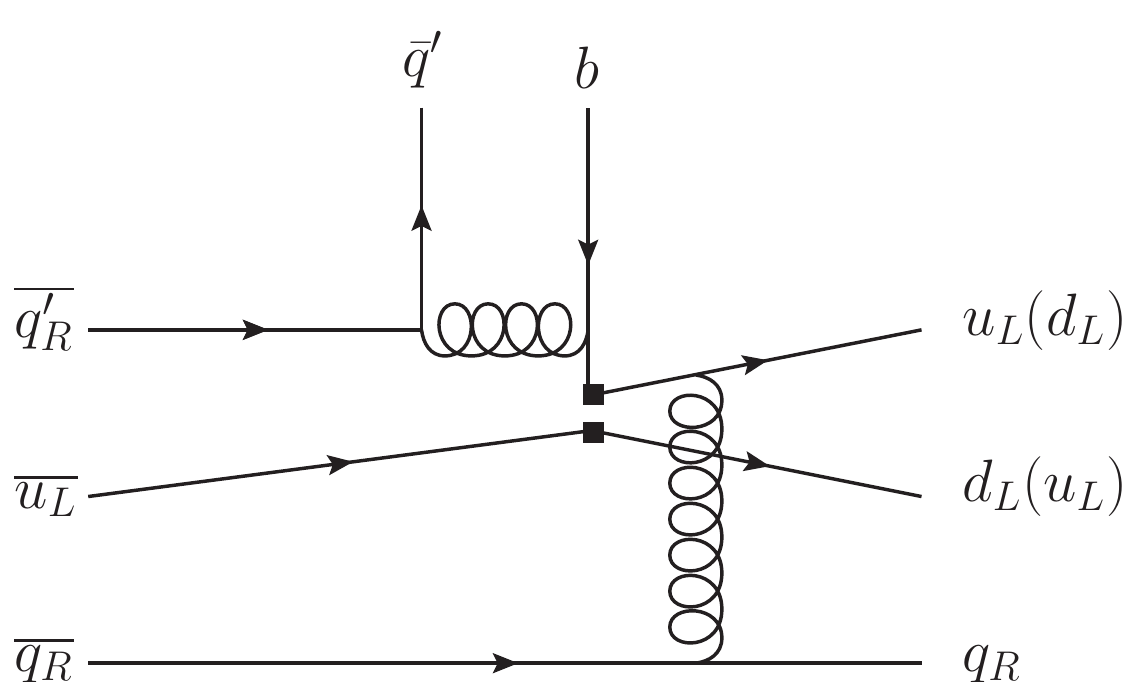}
}\subfigure[]{
     \includegraphics[width=0.33\textwidth,natwidth=610,natheight=642]{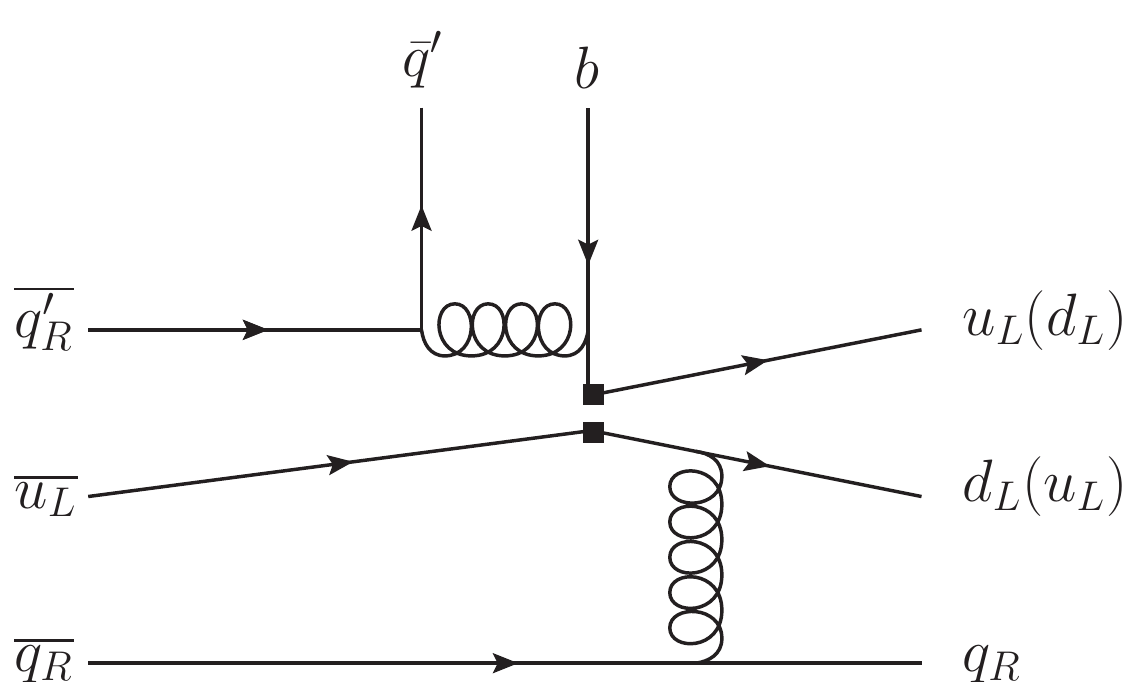}
}\subfigure[]{
     \includegraphics[width=0.33\textwidth,natwidth=610,natheight=642]{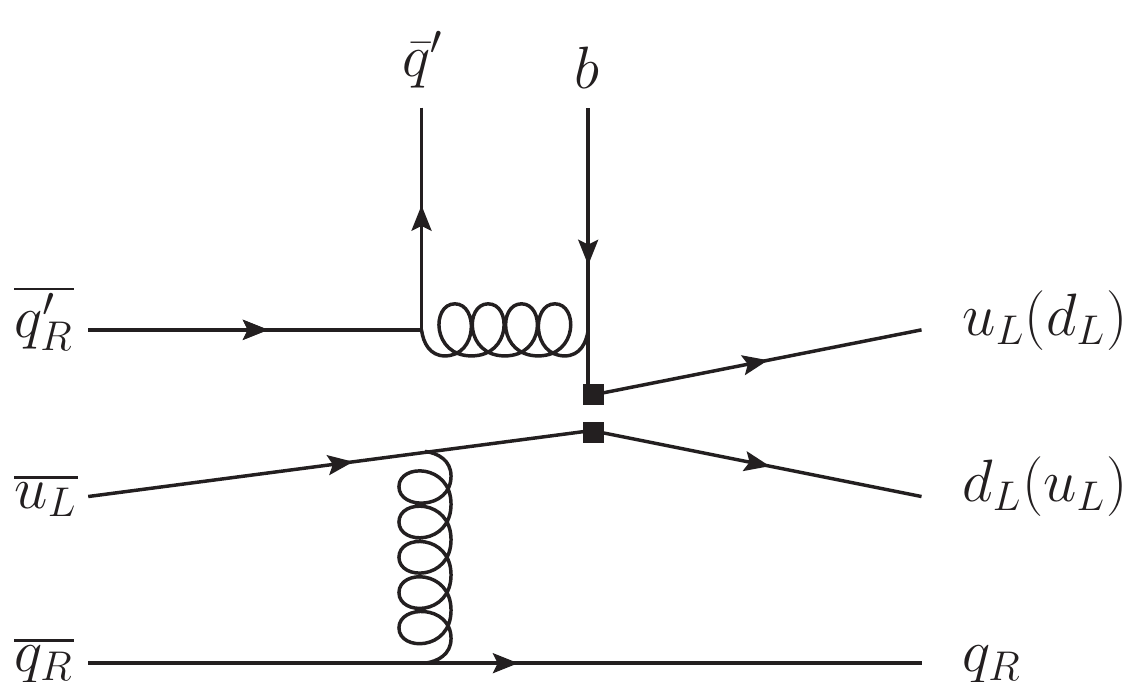}
}
\\
\subfigure[]{
  \includegraphics[width=0.33\textwidth,natwidth=610,natheight=642]{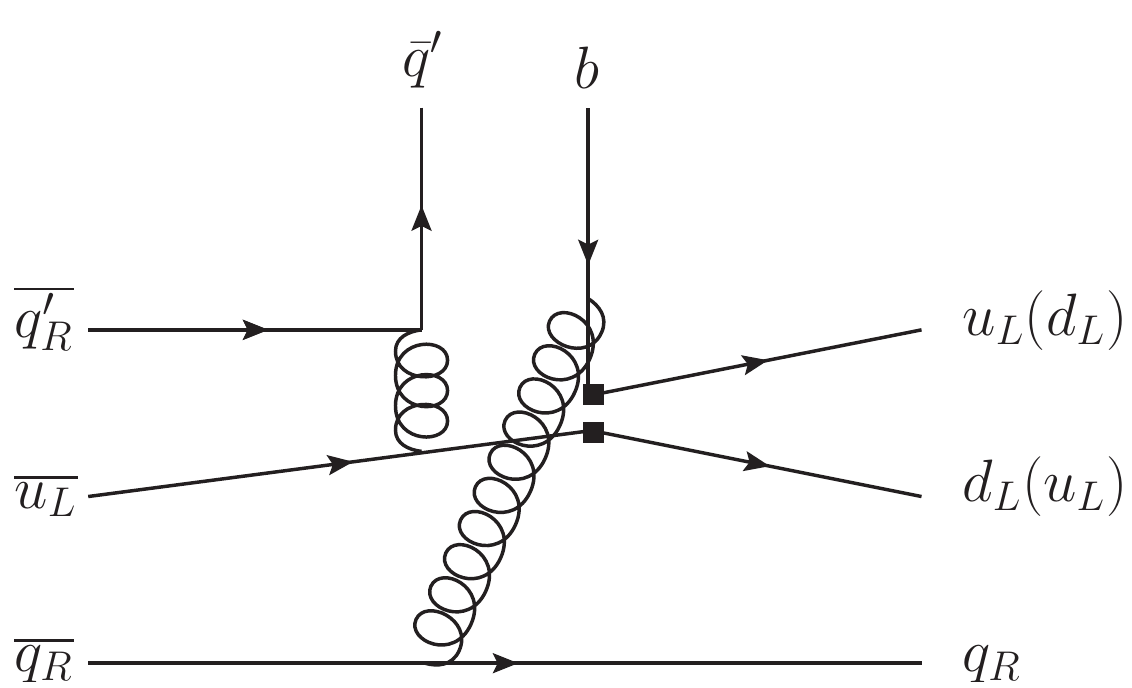}
}\subfigure[]{
  \includegraphics[width=0.33\textwidth,natwidth=610,natheight=642]{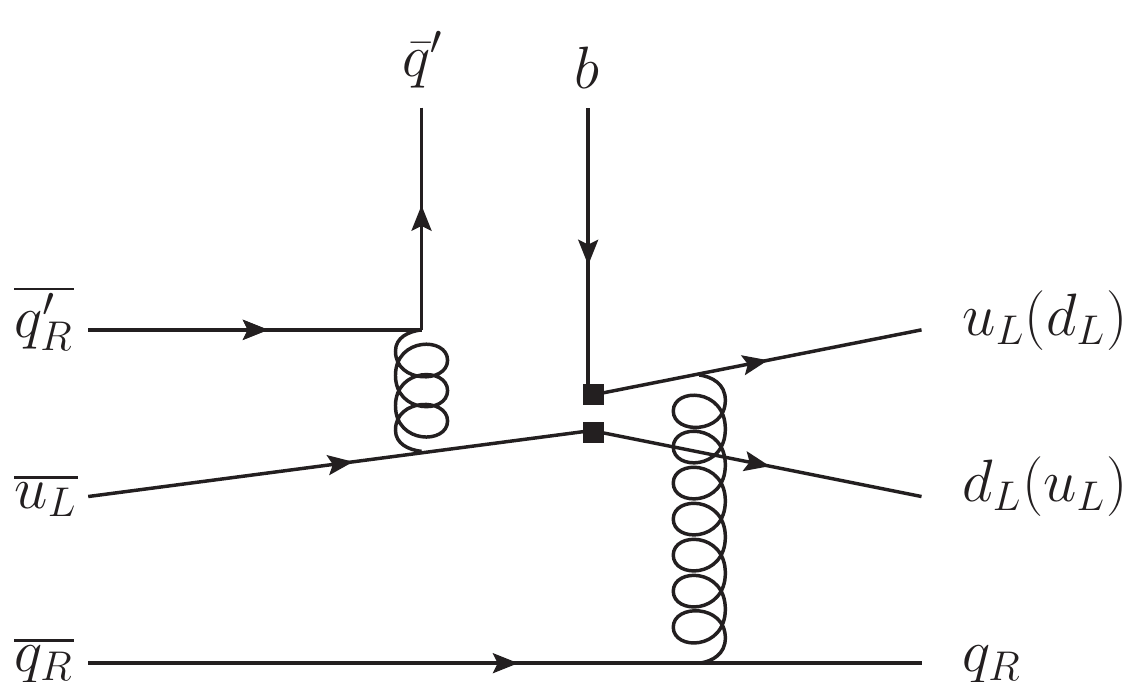}
}\subfigure[]{
  \includegraphics[width=0.33\textwidth,natwidth=610,natheight=642]{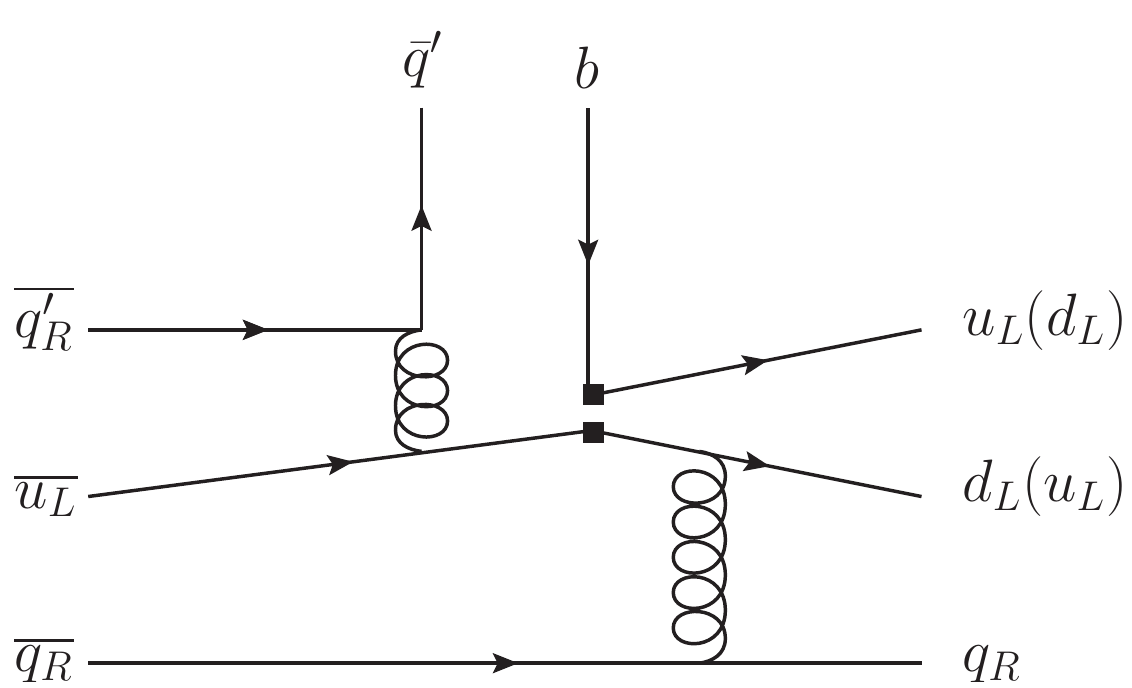}
}
\\
\subfigure[]{
  \includegraphics[width=0.33\textwidth,natwidth=610,natheight=642]{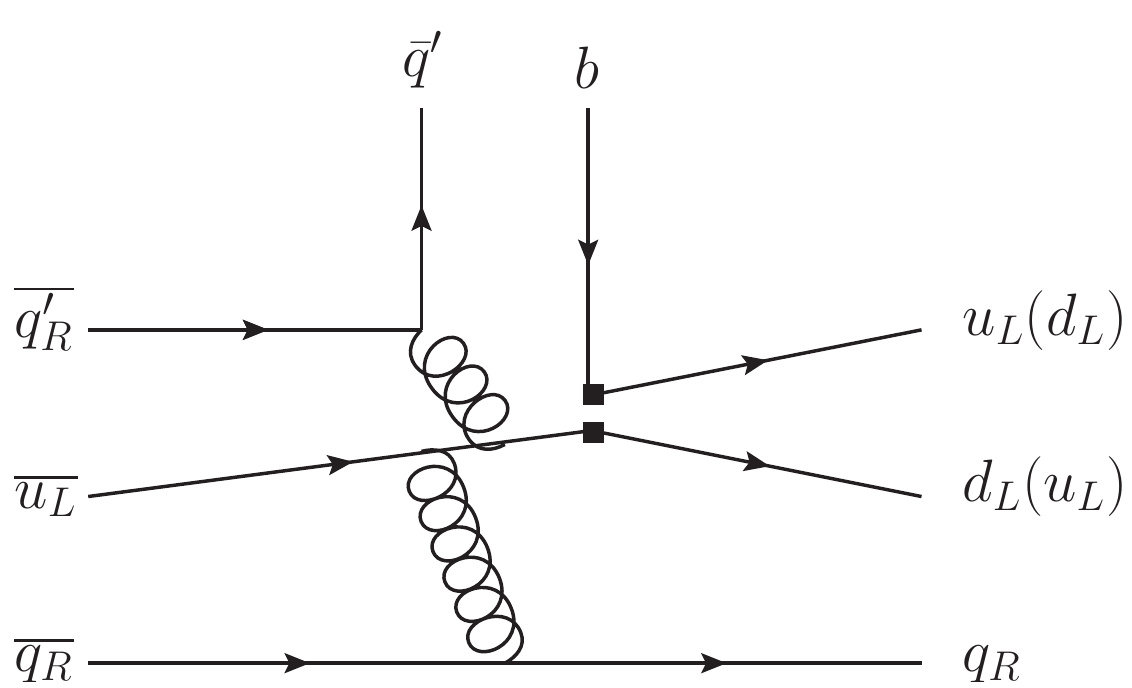}
}\subfigure[]{
  \includegraphics[width=0.33\textwidth,natwidth=610,natheight=642]{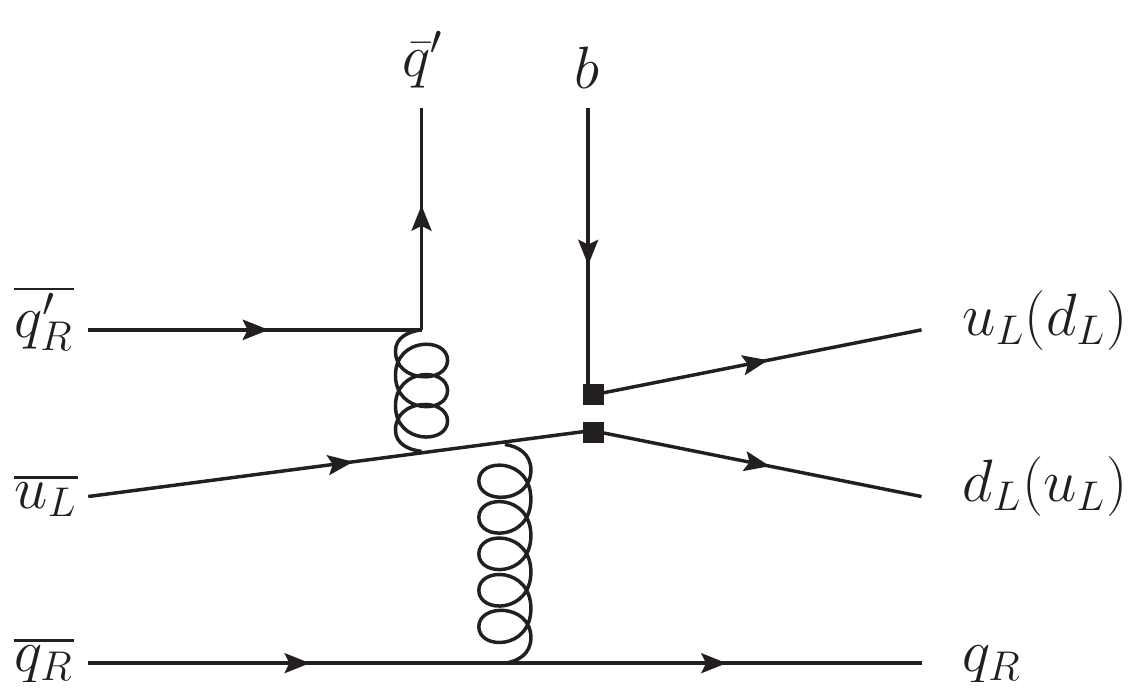}
}\subfigure[]{
  \includegraphics[width=0.33\textwidth,natwidth=610,natheight=642]{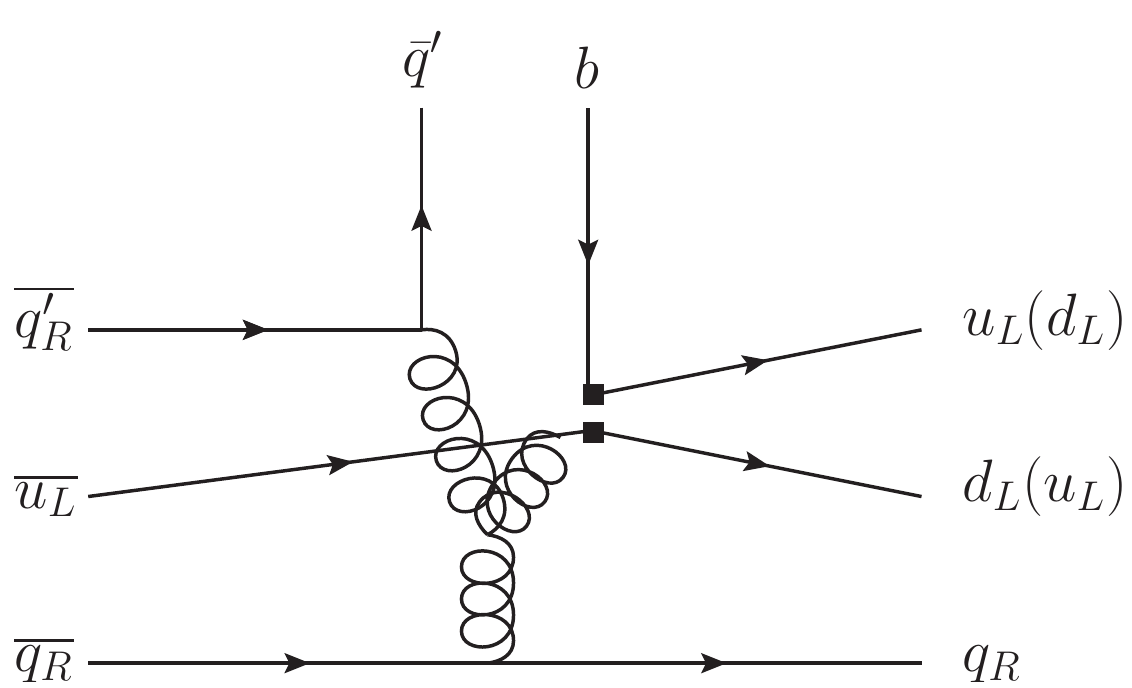}
}
\caption{(a) to (l): Feynman diagrams of internal $W$-emission induced by $O^u_1$. Those in parenthesis are the corresponding diagrams using the Fiertz transformed $O^u_1$ (i.e. $O'{}^u_1$).
These diagrams canceled. 
}
\label{fig:Tgluon cancel}
\end{figure}

\begin{figure}[t!]
\centering
 \subfigure[]{
  \includegraphics[width=0.33\textwidth,natwidth=610,natheight=642]{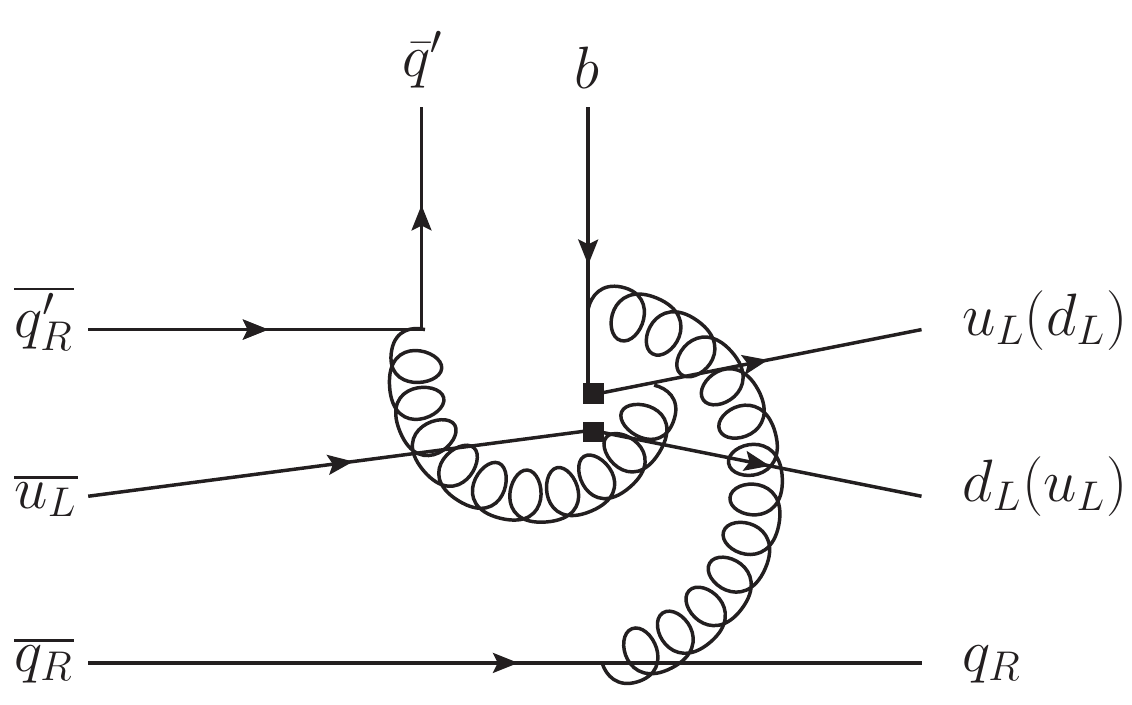}
}\subfigure[]{
  \includegraphics[width=0.33\textwidth,natwidth=610,natheight=642]{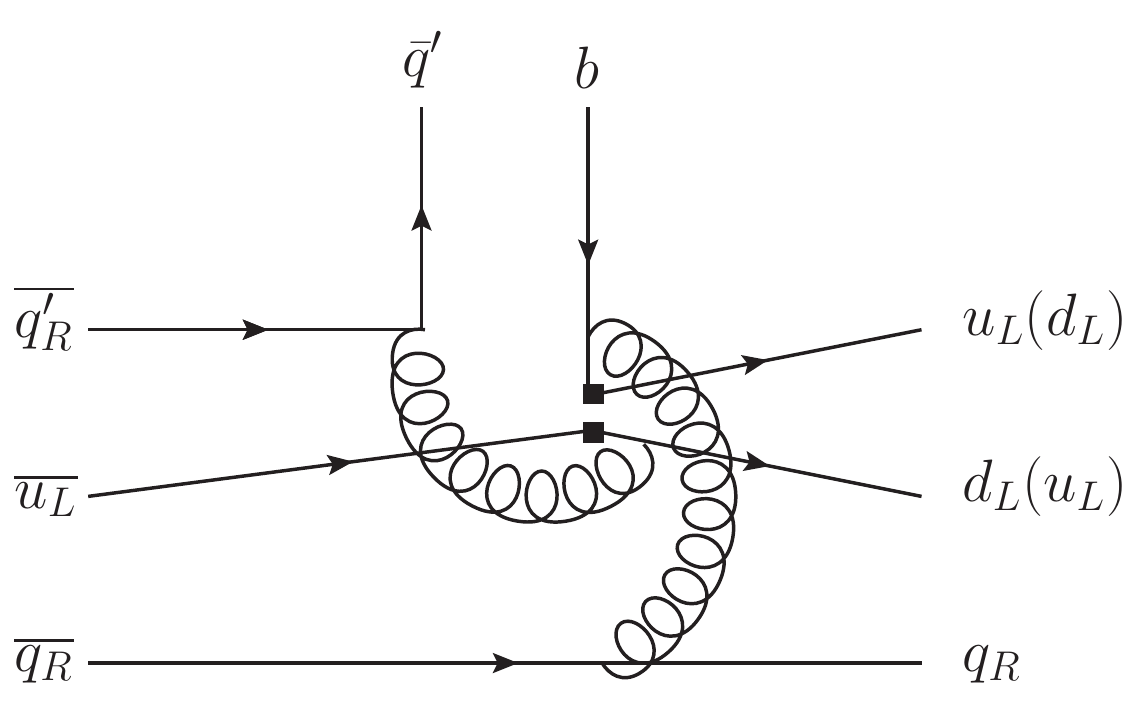}
}\subfigure[]{
  \includegraphics[width=0.33\textwidth,natwidth=610,natheight=642]{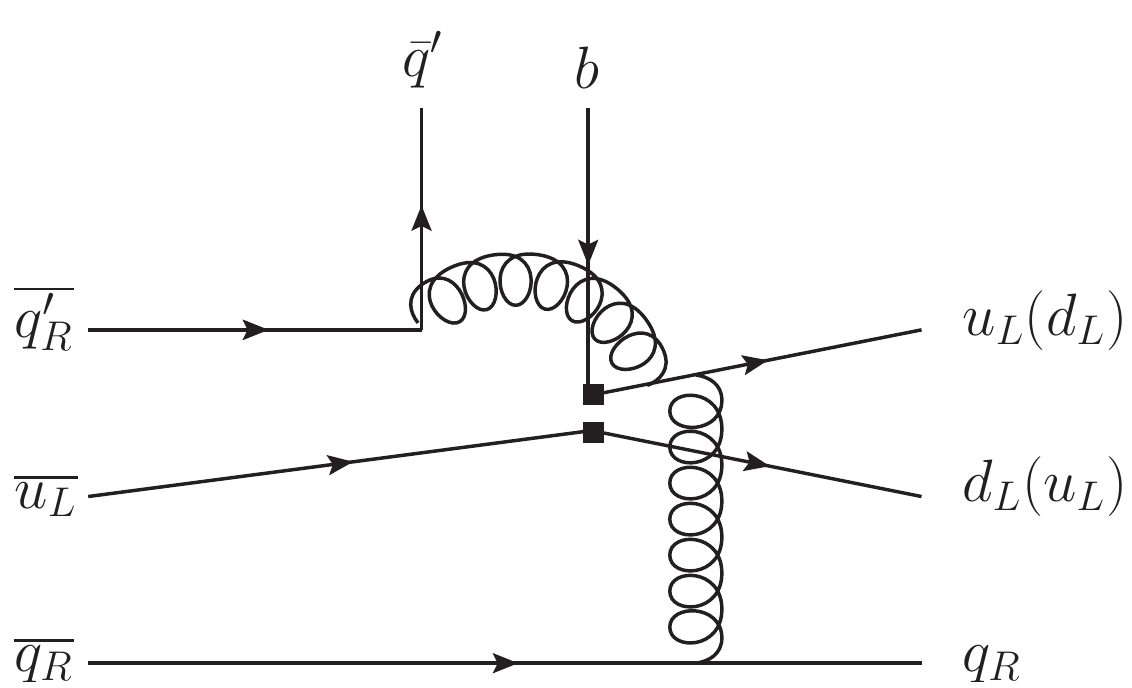}
}
\\
\subfigure[]{
  \includegraphics[width=0.33\textwidth,natwidth=610,natheight=642]{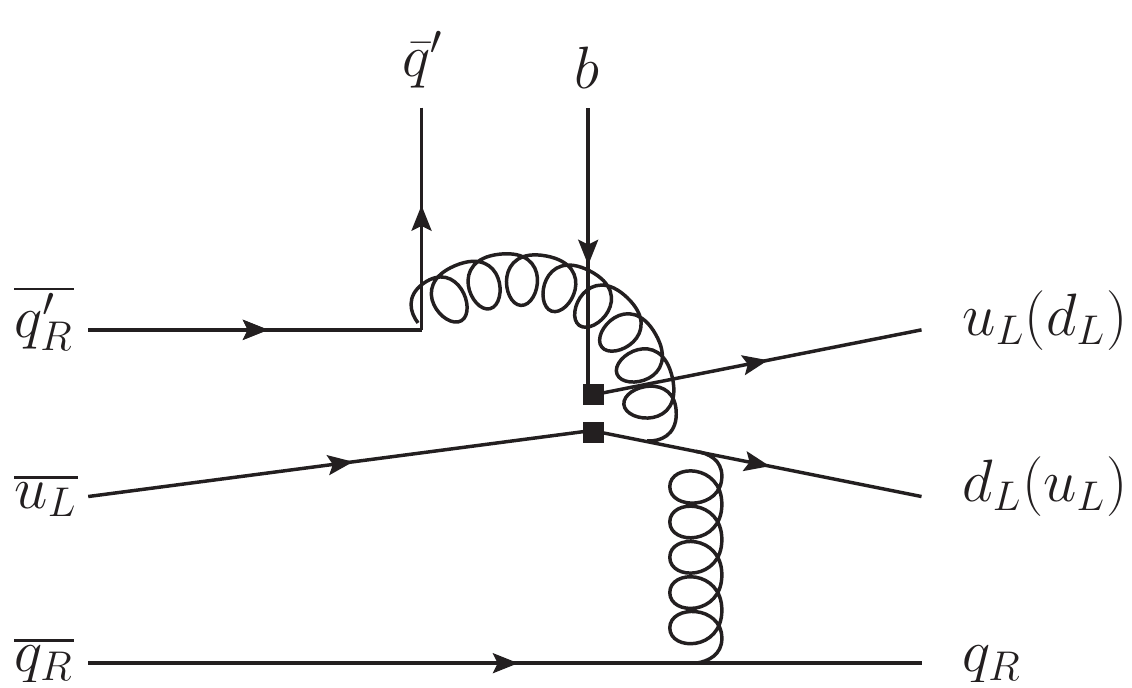}
}\subfigure[]{
  \includegraphics[width=0.33\textwidth,natwidth=610,natheight=642]{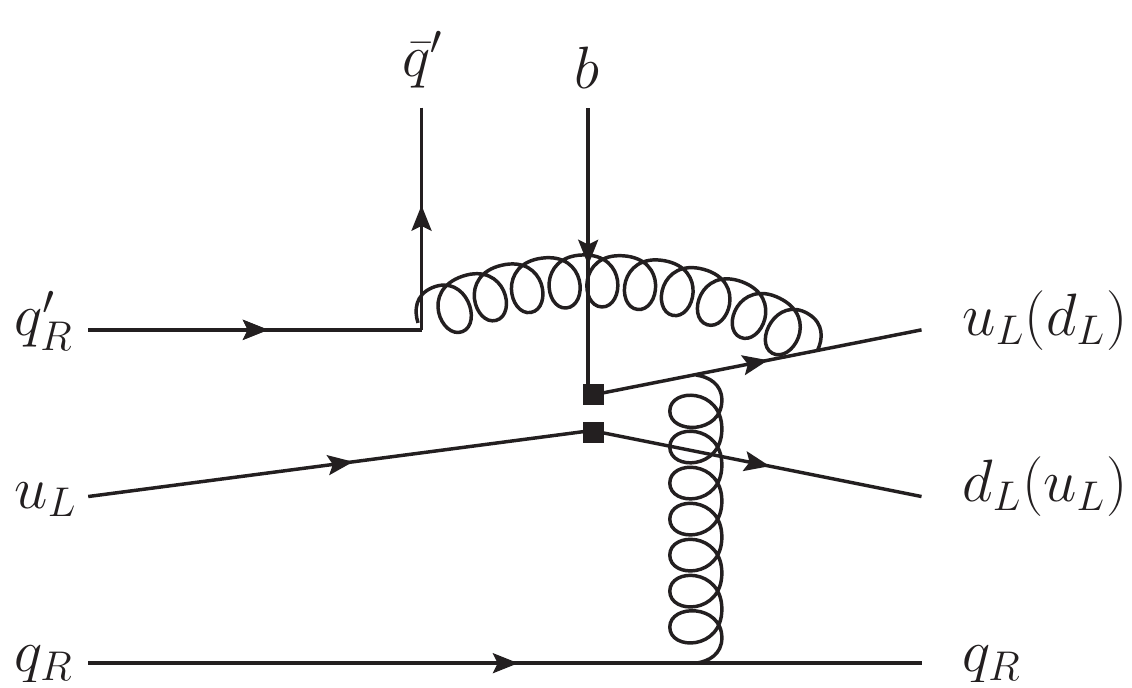}
}\subfigure[]{
  \includegraphics[width=0.33\textwidth,natwidth=610,natheight=642]{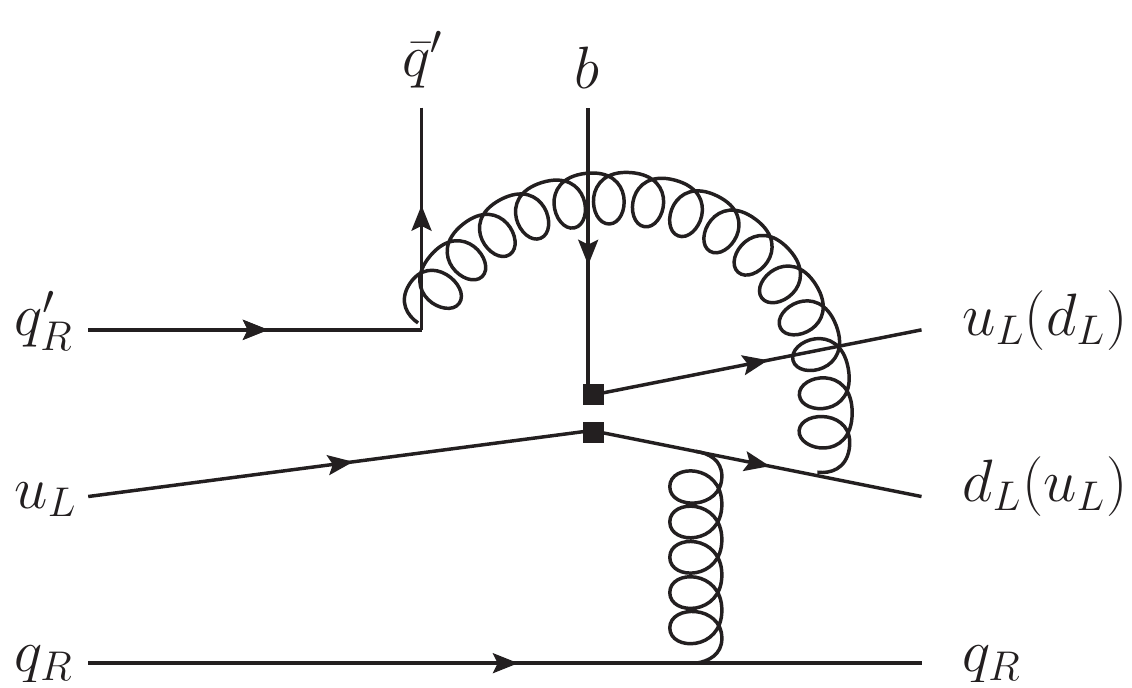}
}
\\
\subfigure[]{
  \includegraphics[width=0.33\textwidth,natwidth=610,natheight=642]{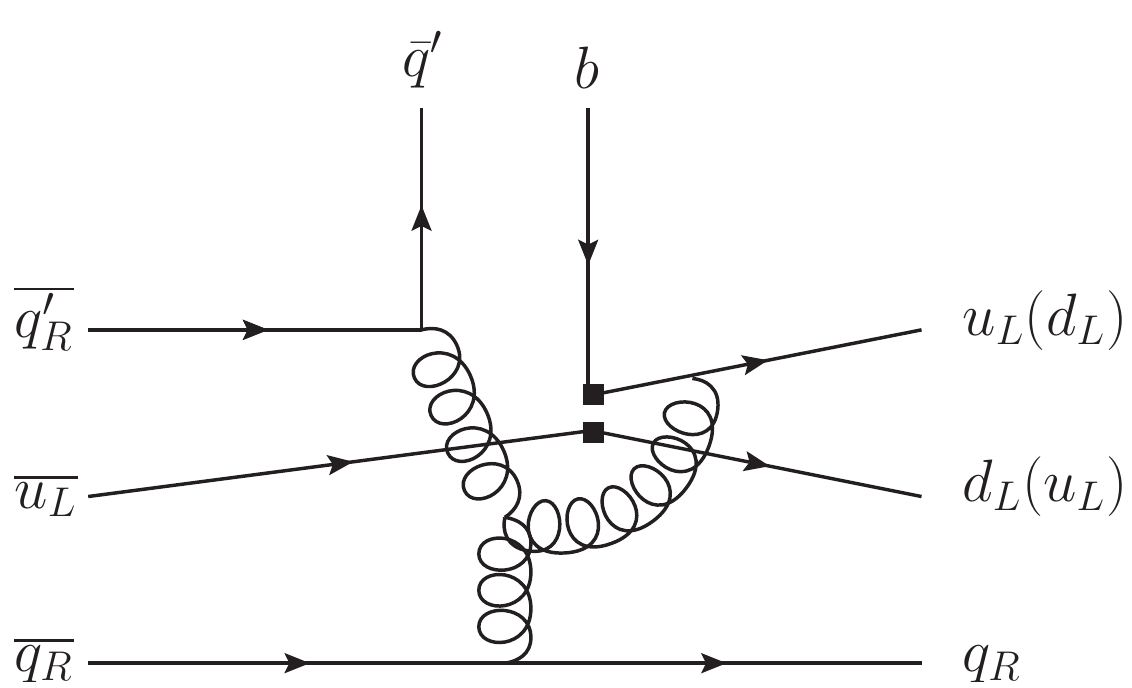}
}\subfigure[]{
  \includegraphics[width=0.33\textwidth,natwidth=610,natheight=642]{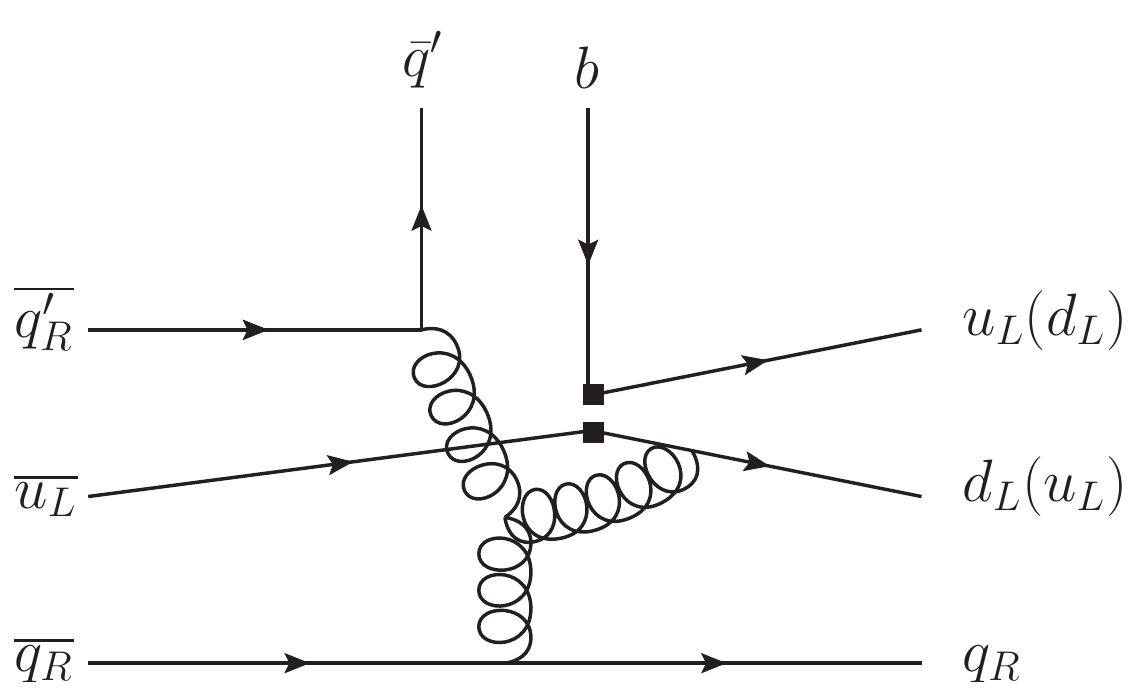}
}\subfigure[]{
  \includegraphics[width=0.33\textwidth,natwidth=610,natheight=642]{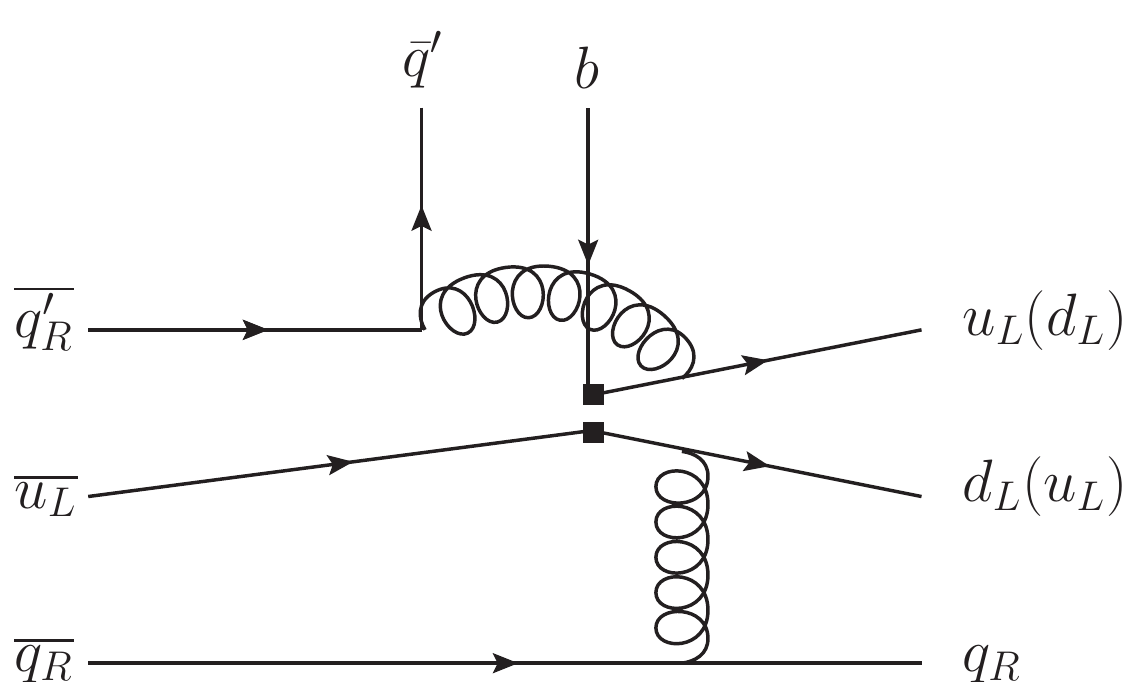}
}
\\
\subfigure[]{
  \includegraphics[width=0.33\textwidth,natwidth=610,natheight=642]{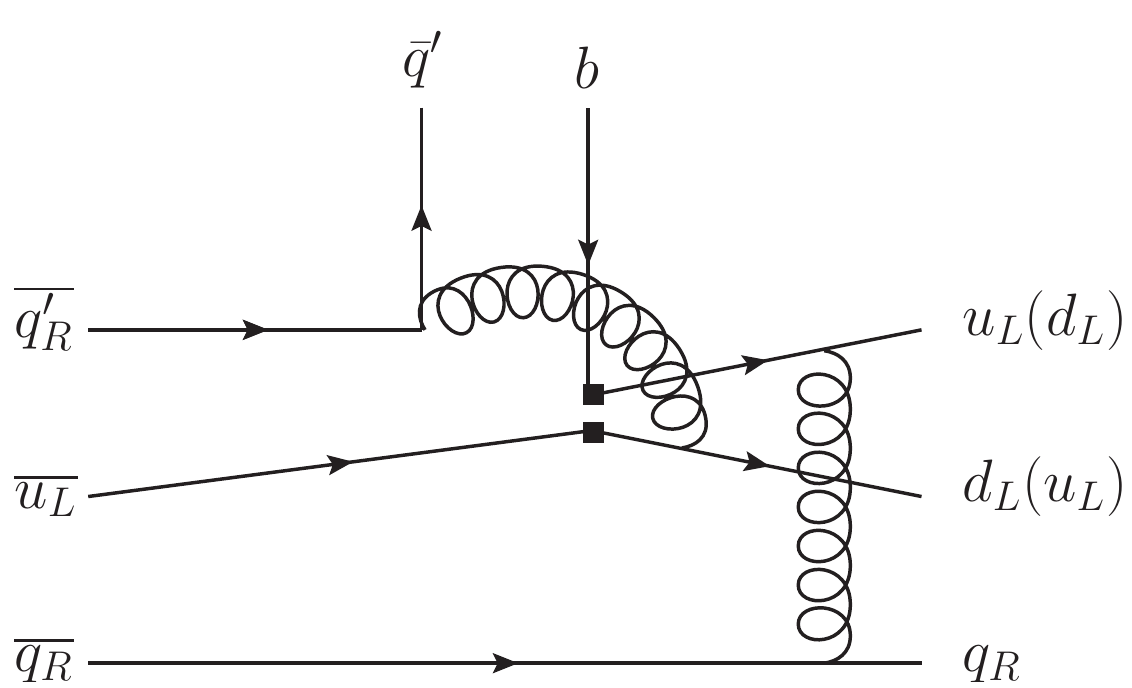}
}\subfigure[]{
  \includegraphics[width=0.33\textwidth,natwidth=610,natheight=642]{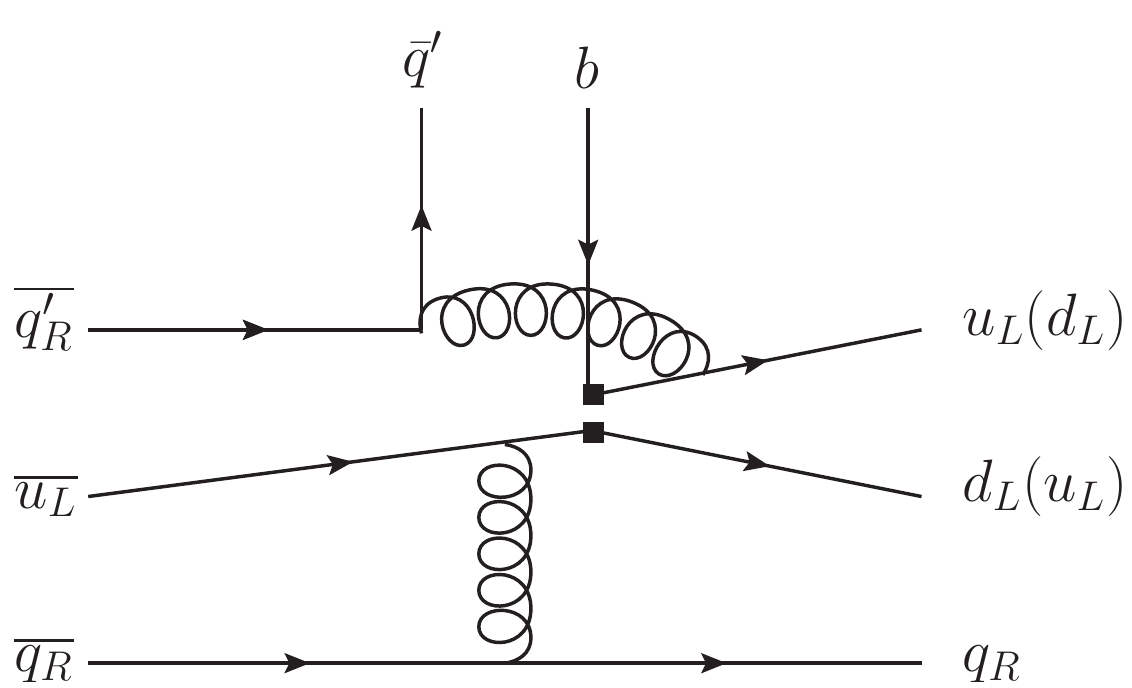}
}\subfigure[]{
  \includegraphics[width=0.33\textwidth,natwidth=610,natheight=642]{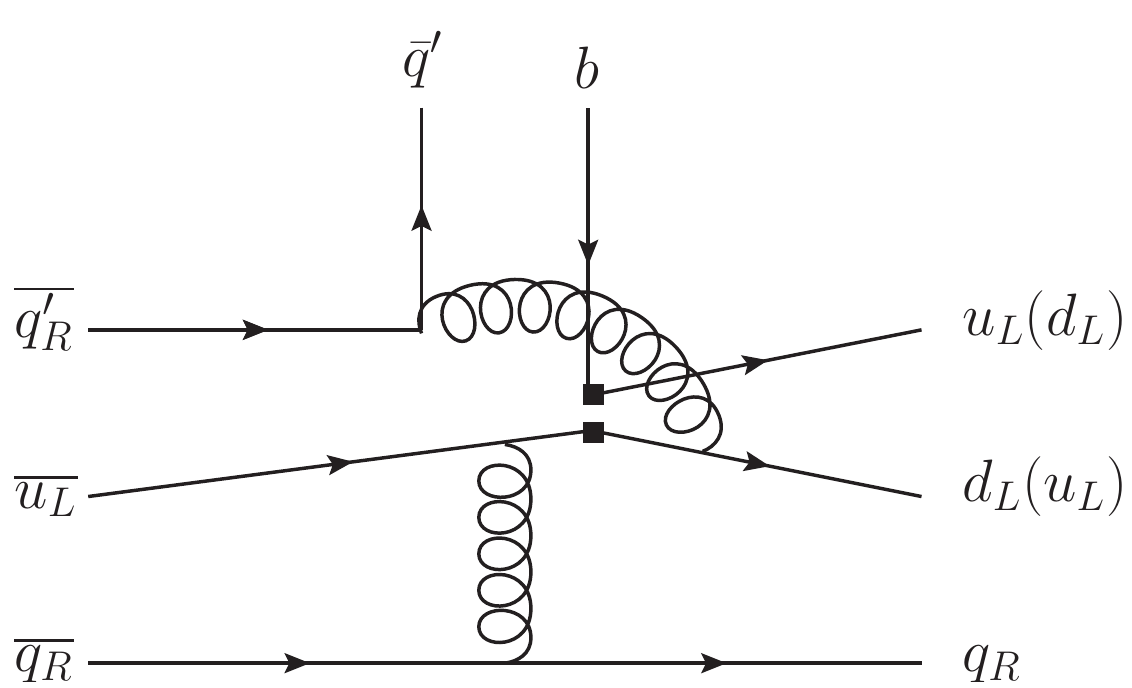}
}
\caption{(a) to (l): Same as Fig. 2, but
these diagrams do not cancel.
}
\label{fig:Tgluon not cancel}
\end{figure}

{\footnotesize
\begin{table}[t!]
\caption{\label{tab:color factor}
Color factors for the diagrams in Figs. 2 and  3.
Second and third columns are the color factors for amplitudes generated through $O^u_1$ and $O'{}^u_1$, respectively.}
\begin{ruledtabular}
\begin{tabular}{lccc}
Configuration
          & $A_{O^u_1}$
          & $A_{O'{}^u_1}$
          & $(A_{O^u_1}+A_{O'{}^u_1})/2$
          \\
\hline
Fig. 2(a)
          & ${2}/{3}$
          & $-{2}/{3}$
          &0
          \\
Fig. 2(b)
          & $-{16}/{3}$
          & ${16}/{3}$
          &0
          \\
Fig. 2(c)
          & $6$
          & $-6$
          & $0$
          \\
Fig. 2(d)
          & $-{16}/{3}$
          & ${16}/{3}$
          &0
          \\
Fig. 2(e)
          & $-{16}/{3}$
          & ${16}/{3}$
          &0
          \\
Fig. 2(f)
          & $-{16}/{3}$
          & ${16}/{3}$
          &0
          \\
Fig. 2(g)
          & ${8}/{3}$
          & $-{8}/{3}$
          & $0$
          \\
Fig. 2(h)
          & ${8}/{3}$
          & $-{8}/{3}$
          & $0$
          \\
Fig. 2(i)
          & ${8}/{3}$
          & $-{8}/{3}$
          & $0$
          \\
Fig. 2(j)
          & ${8}/{3}$
          & $-{8}/{3}$
          & $0$
          \\
Fig. 2(k)
          & ${8}/{3}$
          & $-{8}/{3}$
          & $0$
          \\
Fig. 2(l)
          & $0$
          & $0$
          & $0$
          \\
Fig. 3(a)
          & ${2}/{3}$
          & $-{8}/{3}$
          & $-1$
          \\
Fig. 3(b)
          & ${8}/{3}$
          & $-{2}/{3}$
          & $1$
          \\
Fig. 3(c)
          & $-{16}/{3}$
          & $-{8}/{3}$
          & $-4$
          \\
Fig. 3(d)
          & ${8}/{3}$
          & ${16}/{3}$
          & $4$
          \\
Fig. 3(e)
          & ${2}/{3}$
          & $-{8}/{3}$
          & $-1$
          \\
Fig. 3(f)
          & ${8}/{3}$
          & $-{2}/{3}$
          & $1$
          \\
FIg. 3(g)
          & $6$
          & $0$
          & $3$
          \\
Fig. 3(h)
          & $0$
          & $-6$
          & $-3$
          \\
Fig. 3(i)
          & $-{16}/{3}$
          & $-{8}/{3}$
          & $-4$
          \\
Fig. 3(j)
          & ${16}/{3}$
          & ${8}/{3}$
          & $4$
          \\
Fig. 3(k)
          & $-{16}/{3}$
          & $-{8}/{3}$
          & $-4$
          \\
Fig. 3(l)
          & ${8}/{3}$
          & ${16}/{3}$
          & $4$
          \\
\end{tabular}
\end{ruledtabular}
\end{table}
}

In charmless baryonic $B$ decays, the three quarks of
the energetic light baryon almost share the same momentum fraction $\sim 1/3$. Hence, at least two hard gluons are needed to produce an
energetic light baryon: one hard gluon to kick the spectator quark of
the $B$ meson to make it energetic and the other to produce
the hard $q\bar q$ pair. In the Feynman diagrams shown in Figs. 2 and 3, we add hard gluons explicitly.
For simplicity, we only show the diagrams for $\Delta S=0$ transition. Those for $\Delta S=-1$ can be easily obtained by changing the final state $d_L$ into $s_L$.

We first consider the tree amplitudes generated by the $O^u_1$ operator.
The $O^u_1$ and the Fiertz transformed $O'{}^u_1$ are given by
\be
O^u_1=(\bar u_\alpha b_\alpha)_{V-A}(\bar q_\beta u_\beta)_{V-A},
\quad
O'{}^u_1=(\bar q_\beta b_\alpha)_{V-A}(\bar u_\alpha u_\beta)_{V-A},
\label{eq: O1}
\en
with $q=d,s$. Although $O'{}^u_1$ is identical to $O^u_1$,
we purposely denote it with a different notation for the sake of the ensuing discussion.
To proceed, we replace $O^u_1$ by $O'{}^u_1$ in each of the Feynman diagrams with a suitable replacement of color and flavor indices while keep Dirac and momentum structures intact.
Since $O'{}^u_1$ is equal to $O^u_1$, the sum of all the diagrams generated from $O'{}^u_1$ should be equal to the sum of all the diagrams generated from $O^u_1$. In other words, we have
\be
\la {\cal B}\overline{\cal B}'|O^u_1|\overline B\ra
=\la {\cal B}\overline{\cal B}'|O'{}^u_1|\overline B\ra
=\frac{1}{2}\left(\la {\cal B}\overline{\cal B}'|O^u_1|\overline B\ra
+\la {\cal B}\overline{\cal B}'|O'{}^u_1|\overline B\ra\right)
\label{eq: OO'}
\en
for the tree amplitudes, and we will inspect the cancelation diagram by diagram through the use of the above equation.
Note that the counter diagram (the diagram obtained by replacing $O^u_1$ by $O'{}^u_1$) switches $u_L$ and $d_L, s_L$ and changes colors. Since the baryon wave function is symmetric in $u_L$ and $d_L (s_L)$, and the counter diagram has the same Dirac and momentum structure as the original one, all we need to check is the change in color factors.
Note that Eq. (\ref{eq: OO'}) should be respected by all the model calculations.

In Table I we give explicitly the color factors for each diagram in Figs. 2 and 3. \footnote{In the pQCD approach, Feynman diagrams for the decay $\ov B^0\to\Lambda_c^+\bar p$ are similar to those in Figs. 2 and 3 for $\ov B^0\to p\bar p$ except for a replacement of the $c$ quark by the $u$ quark. The relevant color factors
have been evaluated in \cite{He:2006b}. Our results agree with them.}
For Fig. 2, we have, for example,
\be
2(b): \epsilon_{\alpha\beta\gamma}
(T^i_{\rho\alpha} T^i_{\delta\rho} )
(T^j_{\sigma\delta} T^j_{\kappa\gamma})
\delta_{\lambda\sigma}\delta_{\eta\beta}
\epsilon_{\lambda\eta\kappa}&=&-\frac{16}{3},
\non\\
2(b)': \epsilon_{\alpha\beta\gamma}
( T^i_{\rho\alpha}T^i_{\delta\rho})
(T^j_{\sigma\delta} T^j_{\kappa\gamma})
\delta_{\eta\sigma}\delta_{\lambda\beta}
\epsilon_{\lambda\eta\kappa}&=&\frac{16}{3},
\non\\
2(d): \epsilon_{\alpha\beta\gamma}
( T^i_{\rho\alpha}T^i_{\delta\rho})
(T^j_{\lambda\sigma} T^j_{\kappa\gamma})
\delta_{\sigma\delta}\delta_{\eta\beta}
\epsilon_{\lambda\eta\kappa}&=&-\frac{16}{3},
\non\\
2(d)': \epsilon_{\alpha\beta\gamma}
(T^i_{\rho\alpha}T^i_{\delta\rho} )
(T^j_{\lambda\sigma} T^j_{\kappa\gamma})
\delta_{\eta\delta}\delta_{\sigma\beta}
\epsilon_{\lambda\eta\kappa}&=&\frac{16}{3},
\en
where $T$'s arise from the gluon vertices, $\epsilon$ is from the baryon's color structure, for example, $\epsilon_{\alpha\beta\gamma}$ is from the baryon on the left and $\epsilon_{\lambda\eta\kappa}$ is from the baryon on the right, and the Kronecker delta symbols reflect the color structure of $O^u_1$ or $O'{}^u_1$.
Note that the difference between 2(b) [2(d)] and 2(b)' [2(d)'] is the order of the indices of the two Kronecker $\delta$ factors, corresponding to the color structure of $O^u_1$ and $O'{}^u_1$, respectively. The values of these color factors can be easily worked out using the identity
\be
T^i_{\alpha\beta} T^i_{\gamma\delta}=\frac{1}{2}\left(\delta_{\alpha\delta}\delta_{\gamma\beta}
-\frac{1}{3}\delta_{\alpha\beta} \delta_{\gamma\delta}\right).
\en
It is easily seen that the color factor of Fig. 2(b)' [2(d)'] is opposite to that of Fig. 2(b) [2(d)].
We thus conclude that the amplitudes of Figs. 2(b)' and 2(d)' are opposite in sign to that of Figs. 2(b) and 2(d), respectively. These diagrams cancel each other. This applies to all the diagrams in Fig. 2 (see Table I).

There exist other diagrams related to those discussed so far by crossing two  of the fermion lines while keeping the same gluon line attached. For example, a new diagram can be obtained from Fig. 2(d) by shifting the $u_L$ quark line in such a way that the quark line order reads $d_L,u_L$ and $q_R$. The gluon conected to $u_L$ and $q_R$ is still there.
The amplitudes of the new diagrams are identical to the original ones, since crossing a fermion line gives a minus sign, while changing the color indices gives another minus sign that compensates the sign changed. Since momentum changes are irrelevant as noted in passing, these diagrams also cancel.

We next consider the diagrams in Fig. 3. As we shall see shortly, these figures do not cancel each other.
For example, the color factors of the following diagrams are given by
\be
3(c): \epsilon_{\alpha\beta\gamma}
( T^i_{\rho\alpha}T^i_{\delta\sigma})
(T^j_{\lambda\delta} T^j_{\kappa\gamma})
\delta_{\sigma\rho}\delta_{\eta\beta}
\epsilon_{\lambda\eta\kappa}&=&-\frac{16}{3},
\non\\
3(c)': \epsilon_{\alpha\beta\gamma}
(T^i_{\rho\alpha}T^i_{\delta\sigma} )
(T^j_{\lambda\delta} T^j_{\kappa\gamma})
\delta_{\eta\rho}\delta_{\sigma\beta}
\epsilon_{\lambda\eta\kappa}&=&-\frac{8}{3},
\non\\
3(e): \epsilon_{\alpha\beta\gamma}
(T^i_{\rho\alpha}T^i_{\lambda\delta}  )
(T^j_{\delta\sigma}T^j_{\kappa\gamma})
\delta_{\delta\rho}\delta_{\eta\beta}
\epsilon_{\lambda\eta\kappa}&=&\frac{2}{3},
\non\\
3(e)': \epsilon_{\alpha\beta\gamma}
(T^i_{\rho\alpha}T^i_{\lambda\delta} )
( T^j_{\delta\sigma}T^j_{\kappa\gamma})
\delta_{\eta\rho}\delta_{\sigma\beta}
\epsilon_{\lambda\eta\kappa}&=&-\frac{8}{3},
\en
and
\be
3(g): \epsilon_{\alpha\beta\gamma}
T^i_{\rho\alpha}
if_{ijk} T^k_{\lambda\sigma}
 T^j_{\kappa\gamma}
\delta_{\delta\rho}\delta_{\eta\beta}
\epsilon_{\lambda\eta\kappa}
&=&
\epsilon_{\alpha\beta\gamma} T^i_{\rho\alpha}
(T^i_{\lambda\delta} T^j_{\delta\sigma}-T^j_{\lambda\delta} T^i_{\delta\sigma})
 T^j_{\kappa\gamma}
\delta_{\delta\rho}\delta_{\eta\beta}
\epsilon_{\lambda\eta\kappa}
\non\\
&=&3(e)-3(c)=
6,
\non\\
3(g)': \epsilon_{\alpha\beta\gamma} T^i_{\rho\alpha}
if_{ijk} T^k_{\lambda\sigma}
 T^j_{\kappa\gamma}
\delta_{\eta\rho}\delta_{\sigma\beta}
\epsilon_{\lambda\eta\kappa}
&=&3(e)'-3(c)'=
0.
\en
It is clear that amplitudes in Figs. 3(c), 3(e) and 3(g) do not cancel with that of Figs. 3(c)', 3(e)' and 3(g)', respectively. Hence, diagrams in Fig. 3 do not cancel each other (see Table I). As the previous case, diagrams with crossing fermion lines but with the same gluon lines attached are identical to the original ones.

Color factors for all possible hard gluon pairings are summarized in Table 1. We see that 12 configurations from Fig. 2 yield vanishing results when summing over the diagrams form $O^u_1$ and the counter parts from $O'{}^u_1$, while 12 configurations from Fig. 3 survive and can be grouped into 6 pairs, according to the order shown in the table, with opposite color factors within the pairs. The Dirac structure of the amplitudes within the pairs can be related through the Fiertz transformation
\be
(\gamma_\mu P_L)_{ij}(\gamma^\mu P_L)_{kl}=-(\gamma_\mu P_L)_{il}(\gamma^\mu P_L)_{kj},
\label{eq: Fiertz}
\en
and the interchange of the momentum of $u_L$ and $d_L$. Since the baryon wave function is symmetric in momenta, the color factors are opposite and the above Fiertz transformation gives an additional minus sign, the two amplitudes within the pairs are the same and add together, giving non-vanishing results. We thus conclude that the cancellation is incomplete.

To make the above conclusion in a more concrete manner, we write
\be
\la \BB'|O_1^u|\ov B\ra = \sum_i \la \BB'|O_1^u|\ov B\ra_i\,,
\en
where the superscript $i$ refers to the diagrams Figs. 2(a), 2(b),$\cdots$, 3(a), 3(b),$\cdots$, 3(l) induced from the operator $O_1^u$. Denoting the reduced matrix element of $\la \BB'|O_1^u|\ov B\ra_i$ with the color factor being factored out by $\la\la O\ra\ra_i$ and using the results from Table I, we obtain
\be
\la \BB'|O_1^u|\ov B\ra &=& {2\over 3}\la\la O\ra\ra_{2(a)}-{16\over 3}\la\la O\ra\ra_{2(b)}+\cdots+{8\over 3}\la\la O\ra\ra_{2(k)}  \non \\
&& +{2\over 3}\la\la O\ra\ra_{3(a)}+{8\over 3}\la\la O\ra\ra_{3(b)}+\cdots+{8\over 3}\la\la O\ra\ra_{3(k)}, \non \\
\la \BB'|O'{}_1^u|\ov B\ra &=& -{2\over 3}\la\la O\ra\ra_{2(a)}+{16\over 3}\la\la O\ra\ra_{2(b)}+\cdots-{8\over 3}\la\la O\ra\ra_{2(k)}  \non \\
&& -{8\over 3}\la\la O\ra\ra_{3(a)}-{2\over 3}\la\la O\ra\ra_{3(b)}+\cdots+{16\over 3}\la\la O\ra\ra_{3(k)}.
\label{eq: <O>}
\en
It follows from Eq. (\ref{eq: OO'}) that
\be
\la \BB'|O_1^u|\ov B\ra = \la \BB'|O'{}_1^u|\ov B\ra
&=& -\Big(\la\la O\ra\ra_{3(a)}-\la\la O\ra\ra_{3(b)}\Big)
-4\Big(\la\la O\ra\ra_{3(c)} -\la\la O\ra\ra_{3(d)}\Big)
\non \\
&& +\cdots-4\Big(\la\la O\ra\ra_{3(k)}-\la\la O\ra\ra_{3(l)}\Big)
\non\\
&=& -2\la\la O\ra\ra_{3(a)}-8\la\la O\ra\ra_{3(c)}-2\la\la O\ra\ra_{3(e)} \non \\ && +6\la\la O\ra\ra_{3(g)}-8\la\la O\ra\ra_{3(i)}
-8\la\la O\ra\ra_{3(k)},
\en
where use of Eq.~(\ref{eq: Fiertz}) has been made for the last line.
This shows the complete cancellation from Fig. 2 but not so from Fig. 3.

\section{Discussions}

Thus far we have focused on the tree operator $O_1^u$ and its Fiertz transformed one $O'{}^u_1$. Considering the operator $O_2^u$ and its Fiertz transformed one $O'{}^u_2$
\be
O^u_2=(\bar u_\beta b_\alpha)_{V-A}(\bar q_\alpha u_\beta)_{V-A},
\qquad\qquad
O'{}^u_2=(\bar q_\alpha b_\alpha)_{V-A}(\bar u_\beta u_\beta)_{V-A},
\label{eq: O2}
\en
it is easily seen that $O^u_2$ ($O'{}^u_2$) is identical to $O'{}^u_1$ ($O^u_1$), but with $\overline {q_L}$ and $\overline {u_L}$ interchanged. Since the baryon wave functions are symmetric under the above exchange (i.e. exchanging $d_L(s_L)$ with $u_L$), we are led to
\be
\label{eq: O1O2}
\la \BB'|O_2^u|\ov B\ra = \la \BB'|O'{}_2^u|\ov B\ra=
\la \BB'|O_1^u|\ov B\ra = \la \BB'|O'{}_1^u|\ov B\ra.
\en
As a consequence,
\be
\la \BB'|c_1 O_1^u+c_2 O_2^u|\ov B\ra &=& (c_1+c_2)\Big[-2\la\la O\ra\ra_{3(a)}-8\la\la O\ra\ra_{3(c)}-2\la\la O\ra\ra_{3(e)} \non \\ && +6\la\la O\ra\ra_{3(g)}-8\la\la O\ra\ra_{3(i)}
-8\la\la O\ra\ra_{3(k)}\Big].
\en
This shows that the tree amplitude of the baryonic $B$ decay $\ov B\to\BB'$ is proportional to the Wilson coefficient $c_1+c_2$.

In the literature, it is often argued that the tree amplitude is proportional to $c_1-c_2$ (see e.g. \cite{Cheng:2002,Ball,Chang:2001jt,Korner:1988mx}), whereas our conclusion is the other way around. To clarify this point, we write
\be
c_1 O^u_1+c_2 O^u_2=\frac{c_1-c_2}{2}(O{}^u_1-O^u_2)+\frac{c_1+c_2}{2}(O{}^u_1+O^u_2).
\en
It is easy to check that the first (second) term is antisymmetric (symmetric) in the color indices of the initial ($b$, $u$) and final ($\ov{u_L}$ and $\ov{q_L}$) states. Since the baryon-color wave function is totally antisymmetric, it is tempting to claim that only the color-antitriplet operator $O_1-O_2$ contributes. While this argument holds for the diagrams in Figs. 1 and 2, as one can see by using $\la \BB'|c_1O_1^u+c_2 O_2^u|\ov B\ra=c_1\la \BB'|O_1^u|\ov B\ra+c_2\la \BB'|O'{}_1^u|\ov B\ra$ together with Eq. (\ref{eq: <O>}), it is no longer true for those diagrams in Fig. 3,  where the color structure of the amplitude is affected by the presence of gluon exchanges. Even for Figs. 1 and 2, contributions from the Fiertz re-ordered operators $O'{}^u_1$ and $O'{}^u_2$, which were missed in the literature, should be taken into account. As a result of Eq. (\ref{eq: O1O2}), the tree amplitude induced by the operator $O_1-O_2$ vanishes and this is consistent with our previous argument for the vanishing tree amplitude of Fig. 1\,.

To discuss the  diagrams depicted in Figs. 2 and 3, it is more convenient to write
\be
c_1 O^u_1+c_2 O^u_2=\frac{c_1-c_2}{2}(O'{}^u_1-O^u_2)+\frac{c_1+c_2}{2}(O'{}^u_1+O^u_2).
\en
It is obvious that
the first term is antisymmetric in $\overline {q_L}$ and $\overline {u_L}$ and, most importantly, this feature holds irrespective of the QCD color interaction.  Since the baryon wave function is symmetric in $q_L(=d_L,s_L)$ and $u_L$ and (perturbative) QCD respects chirality and flavor, this term does not contribute to the internal $W$-emission amplitudes. This reinforces our conclusion that the tree amplitude is proportional to the Wilson coefficient combination $c_1+c_2$ rather than $c_1-c_2$.

The above features apply to all $(V-A)\otimes(V-A)$ operators, namely $O_{1,2}^u, O_3, O_4, O_9, O_{10}$, in all $B$ to charmless two-body baryonic decays with low-lying octet and/or decuplet final states.
Therefore, $c_3, c_4, c_9, c_{10}$ appear in the penguin amplitudes also in the form of $c_3+c_4$ and $c_9+c_{10}$.
Things are different in the case of $O_5, O_6, O_7, O_8$. Since their forms will be changed after the Fiertz transformation, the above argument is not applicable. For example, although we can also write
\be
c_5 O_5+c_6 O_6=\frac{c_5-c_6}{2}(O'_5-O_6)+\frac{c_5+c_6}{2}(O'_5+O_6),
\en
the first (second) term is no longer antisymmetric (symmetric) in $\overline {q_L}$ and $\overline {u_L}$ and the previous argument breaks down. The relative sign between $c_5$ and $c_6$ ($c_7$ and $c_8$) cannot be fixed by the symmetry alone.

Note that the above-mentioned partial cancellation (i.e. cancellation from half of the Feynman diagrams) occurs in the tree-dominated charmless mode $\BB'$, but not in the charmful states ${\cal B}_c\ov{\cal B}$ and ${\cal B}_c\ov{\cal B}'{}_c$.
We thus advocate that the partial cancellation is responsible the dynamical suppression $f_{dyn}$ of $\ov B^0\to p\bar p$ relative to $\ov B^0\to\Lambda_c\bar p$ apart from the CKM suppression.
Of course, this conjecture remains to be checked by realistic pQCD calculations.

Finally we would like to comment on the previous model calculations. The internal $W$-emission amplitude for charmless two-body baryonic modes is often expressed as $(c_1-c_2)\la \BB'|O_1^u-O_2^u|\ov B\ra$. When taking into account the contributions from the Fiertz re-ordered operators $O'{}^u_1$ and $O'{}^u_2$, the hadronic matrix element vanishes as one can see from Eq. (\ref{eq: O1O2}). In the pole model, the internal $W$-emission amplitude of, for example, $\ov B^0\to p\bar p$ is proportional to $(c_1-c_2)\la p|O_1^u-O_2^u|\Sigma_b^+\ra=2(c_1-c_2)\la p|O_1^u|\Sigma_b^+\ra$ \cite{Cheng:2002}. However, when the contribution from $O'{}^u_1$ is included, it cancels the one from $O^u_1$ due to the fact that the proton wave function
is symmetric in the flavor and momenta of $u_L$ and $d_L$ but antisymmetric in color indices. As stressed in passing, the internal $W$-emission amplitude should be of the form $(c_1+c_2)\la \BB'|O_1^u+O_2^u|\ov B\ra$ and the pQCD approach will be the most reliable approach to evaluate the relevant hadronic matrix elements.

\section{Conclusions}
Charmless two-body baryonic $B$ decays are very rare. The first mode observed recently by LHCb was $\ov B^0\to p\bar p$ with a branching fraction of order $10^{-8}$.  Tree-dominated charmless baryonic decays such as $\ov B^0\to p\bar p, \Lambda\bar\Lambda$ proceed mainly through the internal $W$-emission diagram. All the earlier model predictions are too large compared with experiment. We point out that for a given tree operator $O_i$, the contribution from its Fiertz transformed operator $O'_i$, an effect missed in the literature, has to be taken into account.
Feynman diagrams responsible for internal $W$-emission can be classified into two categories. We found that diagrams in the first category induced by $O_i$ are completely canceled by that from $O'_i$, while no cancellation occurs for diagrams in the second category. The cancellation is ascribed to the fact that
the wave function of low-lying baryons are symmetric in momenta and the quark flavor with the same chirality, but antisymmetric in color indices.  We advocate that the partial cancellation accounts for the smallness of the tree-dominated charmless two-body baryonic $B$ decays which can be checked by realistic pQCD calculations. A by product of this work is that, contrary to the claim in the literature, the internal $W$-emission tree amplitude should be proportional to the Wilson coefficient combination $c_1+c_2$ rather than $c_1-c_2$.

\section{Acknowledgments}

This research was supported in part by the Ministry of Science and Technology of R.O.C. under Grant
Nos. 103-2112-M-001-005 and  103-2112-M-033-002-MY3, and the National Science
Council of R.O.C. under Grant No. NSC100-2112-M-033-001-MY3.



\end{document}